\documentclass[aps,twocolumn,superscriptaddress]{revtex4-1}     

\usepackage{graphicx}
\usepackage{amsmath}
\usepackage{bm}
\usepackage{color}

\begin{document}
\renewcommand{\vec}[1]{\boldsymbol{#1}}
\newcommand{\up}{{\uparrow}}
\newcommand{\dw}{{\downarrow}}
\newcommand{\pa}{{\partial}}
\newcommand{\pd}{{\phantom{\dagger}}}
\newcommand{\bs}[1]{\boldsymbol{#1}}
\newcommand{\todo}[1]{{\textbf{\color{red}#1}}}
\newcommand{\eps}{{\varepsilon}}
\newcommand{\nn}{\nonumber}
\newcommand{\ie}{{\it i.e.},\ }
\def\eg{\emph{e.g.}\ }
\def\ea{\emph{et al.}}

\title{Stability of Disordered Topological Superconducting Phases in Magnet--Superconductor Hybrid Systems}
\author{Eric Mascot}
\affiliation{Department of Physics, University of Illinois at Chicago, Chicago, IL 60607, USA}
\author{Chaitra Agrahar}
\affiliation{Department of Physics, University of Illinois at Chicago, Chicago, IL 60607, USA}
\author{Stephan Rachel}
\affiliation{School of Physics, University of Melbourne, Parkville, VIC 3010, Australia}
\author{Dirk K. Morr}
\affiliation{Department of Physics, University of Illinois at Chicago, Chicago, IL 60607, USA}

\date{\today}

\begin{abstract}
Magnet-superconductor hybrid heterostructures constitute a promising candidate system for the quantum engineering of chiral topological superconductivity. Here, we investigate the stability of their topological phases in the presence of various types of potential and magnetic disorder. In particular we consider magnetic disorder in the coupling strength and spin-orientation, as well as percolation type disorder representing missing magnetic moments. We show that potential disorder leads to the weakest suppression of topological phases, while percolation disorder leads to their strongest suppression. In addition, we demonstrate that in the case of correlated potential disorder, the spatial structure of the disorder potential is correlated not only with the particle number density, but also the Chern number density. Finally, we demonstrate how the disorder-induced destruction of topological superconductivity is reflected in the spatial structure and distribution of the Chern number density.

\end{abstract}

\maketitle

\section{Introduction}

Topological superconductors have attracted much attention in recent years as they represent novel platforms to realize and control Majorana zero modes whose exotic non-Abelian braiding statistics can be employed for the creation of fault-tolerant topological quantum bits \cite{Nayak2010, Sarma2015}. Odd-parity spin-triplet superconductors and, in particular, those that possess a chiral $p$-wave superconducting symmetry \,\cite{read-00prb10267} represent potential realizations of topological superconductors. Candidate materials which might feature such spin-triplet pairing are Sr$_2$RuO$_4$\,\cite{maeno-94n532,mackenzie-03rmp657}, UPt$_3$\,\cite{joynt-02rmp235}, Cu$_x$Bi$_2$Se$_3$\,\cite{hor-10prl057001,sasaki-11prl217001,matano-16np852} and superfluid $^3$He\,\cite{Volovik03}. However, most of these materials have been controversially debated\,\cite{kallin-16rpp054502,sato-17rpp076501,mackenzie-17npjqm40} and the presence of spin-triplet pairing remains to be unambiguously proven.

In addition to these {\it intrinsic} topological superconductors, there has been growing interest in {\it artificial} or {\it engineered} topological superconductors, allowing for the realization of the Kitaev chain\,\cite{kiatev01pu131}, the prototype of a 1D topological superconductor, by proximity-inducing $s$-wave superconductivity\,\cite{fu-08prl096407} in Rashba nanowires\,\cite{lutchyn-10prl077001,oreg-10prl177002,mourik-12s1003}. An alternative approach to the creation of Kitaev chains has been taken with magnet--superconductor hybrid (MSH) structures in which (Shiba) chains of magnetic atoms, either via self-assembly \,\cite{nadj-perge-13prb020407,nadj-perge-14s602,ruby-15prl197204,pawlak-16npjqi16035} or via atomic manipulation techniques \,\cite{kim-18sa5251}, were placed on the surface of s-wave superconductors. These studies were subsequently extended into two dimensions \,\cite{Li2016,rontynen-15prl236803,Rachel2017} through the creation of magnetic Shiba islands in Pb/Co/Si(111)\,\cite{Menard2016} and Fe/Re(0001)-O(2$\times$1)\,\cite{palacio18arXiv1809.04503} heterostructures.

One of the defining properties of topological states of matter is their {\it topological protection} against small perturbations and disorder. On the other hand, the experimental growth of topological materials often leads to significant amounts of disorder, raising the question of what extent of disorder can destroy topological phases. Classical work on disorder effects in intrinsic topological superconductors mainly focused on their bulk properties using continuum Dirac theories \,\cite{ludwig-94prb7526,senthil-00prb9690,foster-14prb155140,evers-08rmp1355}. More recent studies investigated the stability of the topological surface states in topological superconductors\,\cite{queiroz-14prb054501,queiroz-15prb014202} or the emergence of Majorana bound states through random-field disorder \cite{Zhou2017}. In engineered one-dimensional superconductors, impurities and disorder play a particular important role \cite{Brouwer2011,Ren2018} because the experimental evidence often relies on the observation of a zero-bias peak associated with the presence of a Majorana bound state\,\cite{mourik-12s1003,nadj-perge-14s602}. Such a zero-bias peak can also be induced by an impurity \,\cite{bagrets-12prl227005} emphasizing the importance of understanding the effects of disorder on these systems.

MSH heterostructures are particularly suited for the study of disorder effects as disorder can be visualized through scanning tunneling spectroscopy (STS) techniques, which provide simultaneous insight into the topography and spectroscopic (electronic) properties of the constituent magnetic and superconducting subsystems. Indeed, STS experiments measuring the spin-resolved differential conductance have provided evidence for the non-collinear spin structure of Shiba chains \,\cite{kim-18sa5251}, while topography scans have revealed the extent of edge disorder in Shiba islands \,\cite{palacio18arXiv1809.04503}. The question thus naturally arises not only of how disorder affects the topological phase diagram of two-dimensional MSH structures, but also how disorder destroys topological superconducting phases on the microscopic or spatially local level. In this article, we will study these questions by considering the effects of various types of potential and magnetic disorder, and by investigating the spatial correlations between the disorder potential, the particle density, and the Chern number density, and their relation to the macroscopic topologically invariant of the system, the Chern number.

The paper is organized as follows. In Sec.\,II we introduce the theoretical model to describe two-dimensional MSH structures, and discuss how the topological phase diagram in the presence of disorder can be computed by using the real space Chern number. In Sec.\,III we define various types of potential and magnetic disorder, and discuss their effects on the topological phase diagram. In Sec.\,IV we consider correlated potential disorder, and discuss the spatial correlations between disorder potential, particle density, and Chern number density, and their relation to the macroscopic Chern number. In Sec.\,V, we present our conclusions.

\section{Theoretical Model}

We study the effects of disorder on the topological phase diagram of a two-dimensional MSH structure, also referred to as a Shiba lattice. They are created by placing magnetic adatoms on the surface of a conventional $s$-wave superconductor possessing a Rashba spin-orbit interaction on the surface. The Hamiltonian describing a clean (i.e., non-disordered) Shiba lattice is then given by \cite{Li2016}:
\begin{align}\nn
&H = -t \sum_{{\bf r}, {\bm \delta}} \psi_{\bf r}^\dagger \tau_z \otimes \sigma_0 \psi_{{\bf r}+{\bm \delta}}-\mu \sum_{\bf r} \psi_{\bf r}^\dagger \tau_z \otimes \sigma_0 \psi_{\bf r}  \\ \label{eq:Ham}
&~+ i \alpha \sum_{{\bf r}, {\bm \delta} }  \psi_{\bf r}^\dagger \tau_z \otimes \left[ (\vec{\sigma}\times \vec{\delta}) \cdot \hat{z} \right] \psi_{{\bf r}+{\bm \delta}} + \Delta \sum_{\bf r} \psi_{\bf r}^\dagger \tau_x \otimes \sigma_0 \psi_{\bf r} \\
&~+ J \sum_{\bf r} \psi_{\bf r}^\dagger \tau_0 \otimes \left( \vec{S} \cdot \vec{\sigma} \right) \psi_{\bf r} \nn
\end{align}
where $t$ is the hopping parameter between nearest neighbor sites on a square lattice, ${\bm \delta}$ is the vector connecting nearest neighbor sites, $\mu$ is the chemical potential, $\alpha$ is the Rashba spin-orbit coupling, $\Delta$ is the superconducting order parameter, and $\vec{\sigma}$ and $\vec{\tau}$ are vectors of Pauli matrices corresponding to spin and Nambu space, respectively. We use the Nambu spinor $ \psi_{\bf r} = (\psi_{\bf r \uparrow},\psi_{\bf r \downarrow},\psi_{\bf r \downarrow}^\dagger,-\psi_{\bf r \uparrow}^\dagger)^T$ where $\psi_{\bf r \sigma}^\dagger$ ($\psi_{\bf r \sigma}$) creates (annihilates) an electron at site $\bf r$ and spin $\sigma$. The presence of a hard superconducting $s$-wave gap suppresses the Kondo screening of the magnetic adatoms which allows us to treat the spins classically. For the clean system, we assume a ferromagnetic alignment of all spins along the ${\hat z}$-direction, and therefore set $\vec{S}=S(0,0,1)$ with $S$ being the spin's magnitude. Moreover, due to the particle-hole symmetry of the superconducting state, and the broken time-reversal symmetry arising from the presence of magnetic moments, the topological superconductor belongs to class D \cite{Ryu2010}.

To characterize the topological state of the system even in the presence of disorder, which breaks the translational invariance of the system, we compute the topological invariant --  the Chern number \cite{Thouless1982} -- in real space using \cite{Prodan2010a,prodan17,prodan11jpa113001}
\begin{equation} \label{C real space}
C = \frac{1}{2\pi i} \mathrm{Tr} \left[ P [\delta_1 P, \delta_2 P] \right]
\end{equation}
\begin{equation}
\delta_i P = \sum_{m=-Q}^Q c_m e^{-2\pi i m \hat{x}_i / N} P e^{2\pi i m \hat{x}_i / N}
\end{equation}
where $P$ is the projector onto the occupied spectrum in real space, $N^2$ are the number of sites in the system, and $c_m$ are central finite difference coefficients for approximating the partial derivatives.
The coefficients for positive $m$ can be calculated by solving the following linear set of equations for $\vec{c} = (c_1, \dots, c_Q)$:
\begin{equation}
\hat{A} \vec{c} = \vec{b}, \; A_{ij} = 2j^{2i-1}, \; b_i = \delta_{i,1}, \; i,j \in \{1, \dots, Q \}
\end{equation}
while for negative $m$, we have $c_{-m} = -c_m$. To achieve a small error in the calculation of the Chern number, we take the largest possible value of $Q$ given by $Q=N/2$.
As we show below, important insight into the effects of disorder on the stability of a topological superconductor can be gained by considering the scaled Chern number density, defined as the partial trace over spin and Nambu space, and given by
\begin{equation} \label{C density}
C({\bf r}) = \frac{N^2}{2\pi i} \mathrm{Tr}_{\tau,\sigma} \left[ P [\delta_1 P, \delta_2 P] \right]_{\bf r,r}
\end{equation}
such that $C=\sum_{\bf r}[C({\bf r})]/N^2$.

\section{Random Potential and Magnetic Disorder}

To investigate the effects of disorder on the stability of the topological phases, we consider several types of random potential and magnetic disorder.
The random potential disorder is described by the Hamiltonian
\begin{align}
H_U = \sum_{\bf r}  U_{\bf r}  \, \psi_{\bf r}^\dagger \tau_z \otimes \sigma_0  \psi_{\bf r}
\label{eq:pot}
\end{align}
with $U_{\bf r} \in [-w_U, w_U]$ being random variables from a uniform probability distribution in the range from $-w_U$ to $w_U$. Thus, $w_U$ is a measure for the strength of the disorder.

In addition, we consider three different types of magnetic disorder. In the first type, the strength of the magnetic coupling $J$ is disordered, while the spins are still  ferromagnetically aligned along the $z$-axis, as described by the Hamiltonian
\begin{align}
H^{(1)}_J = \sum_{\bf r}  J_{\bf r} S  \ \psi_{\bf r}^\dagger \tau_0 \otimes \sigma_z  \psi_{\bf r}
\end{align}
with $J_{\bf r}S \in [-w_J, w_J]$ being random variables from a uniform probability distribution. The second type of magnetic disorder is one in which the magnetic coupling $J$ is spatially constant, but the direction of the magnetic moments deviates from the $z$-axis. To describe this type of disorder, we replace the magnetic term in the Hamiltonian, Eq.(\ref{eq:Ham}), by
\begin{align}
H^{(2)}_J = J \sum_{\bf r} \psi_{\bf r}^\dagger  \tau_0 \otimes \left( \vec{S}_{\bf r} \cdot \vec{\sigma} \right) \psi_{\bf r}
\end{align}
where $\vec{S}_{\bf r}$ are spins with random directions which are chosen from a uniform distribution over the spherical cap formed by the polar angle $\theta$. That is, $\vec{S}_{\bf r} = S(\sin{\theta_{\bf r}}\cos{\phi_{\bf r}}, \sin{\theta_{\bf r}}\sin{\phi_{\bf r}}, \cos{\theta_{\bf r}})$ where $\phi_{\bf r} \in [-\pi,\pi]$, $\theta_{\bf r} \in [0,\theta]$, and $\theta$ thus reflects the extent of the orientational disorder.  Finally, the third type of magnetic disorder is of the percolation type, in which the moments are aligned along the $z$-axis with uniform $J$, but at each site, there is a probability $p$ that a magnetic moment is missing. We model such a percolation disorder by replacing the magnetic term in the Hamiltonian, Eq.(\ref{eq:Ham}), by
\begin{align}
H^{(3)}_J = JS \sum_{\bf r} \Theta(w_{\bf r} - p) \psi_{\bf r}^\dagger  \tau_0 \otimes \sigma_z \psi_{\bf r}
\end{align}
with random variables $w_{\bf r} \in [0,1]$ and $p$ describing the degree of percolation, i.e., the mean density of missing magnetic adatoms.
\begin{figure}[t]
\begin{center}
\includegraphics[width=8.5cm]{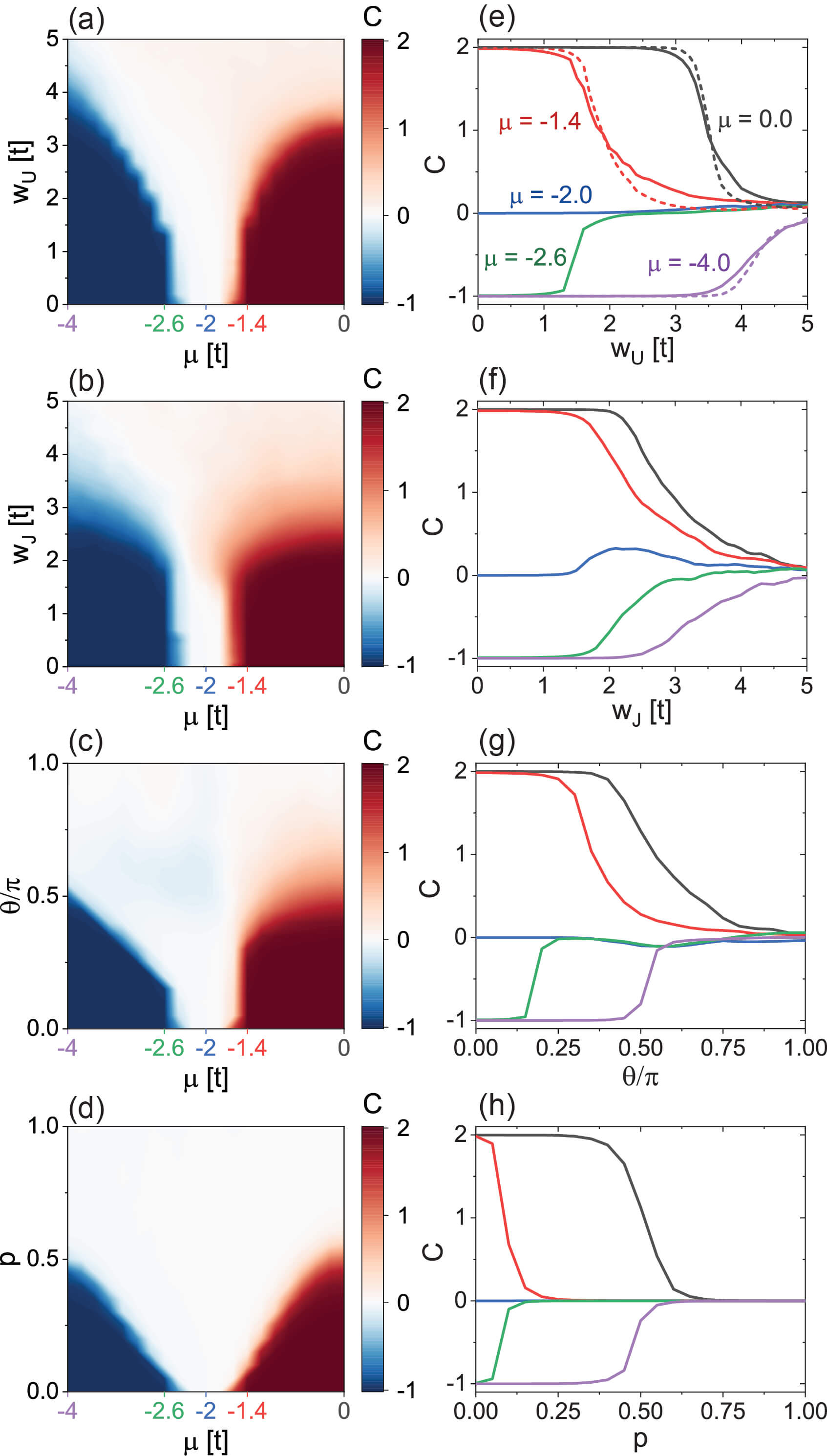}
\caption{Topological phase diagram showing the Chern number as a function of chemical potential, $\mu$, and disorder strength $w$ (as characterized by $w_U, w_J, \theta$ and $p$) for (a),(e) potential disorder, described by $H_U$, and magnetic disorder described by (b),(f) $H^{(1)}_J$, (c),(g) $H^{(2)}_J$, and (d),(h) $H^{(3)}_J$. (e) - (h) solids lines are line cuts of $C$ in panels (a)-(d) as a function of disorder strength for chemical potentials $\mu/t = 0, -1.4, -2.0, -2.6, -4$. The phase diagrams were computed for a $(30 \times 30)$ system with parameters $(JS,\alpha,\Delta)=(2,0.8,1.2)t$. Dashed lines in (e) were obtained from a $(50 \times 50)$ system. }
\label{fig:Chern number}
\end{center}
\end{figure}

All topological phase diagrams shown below are averaged over $n$ disorder realizations. For each type of disorder, $n$ is determined as the smallest number of disorder realization for which the following criterion
\begin{align}
\frac{\sigma(w,\mu)}{\sqrt{n}} < 0.05
\end{align}
is satisfied for disorder strength $w$ (characterized by $w_U, w_J, \theta$ and $p$) and chemical potential $\mu$.
Here, $\sigma$ is the standard deviation of the Chern number for a specific $w$ and $\mu$, and for the results shown below, $10<n<200$.

In Figs.~\ref{fig:Chern number} (a)-(d), we present the topological Chern number phase diagrams for these four types of disorder as a function of chemical potential $\mu$ and disorder strength for a $(30 \times 30)$ site system. For each disorder realization, the Chern number is computed using Eq.(\ref{C real space}). We begin by noting that for a clean system (which corresponds to the zero disorder line in all four phase diagrams) in which all spins are ferromagnetically aligned along the $z$-direction,
the Shiba lattice possesses two topological non-trivial phases with Chern number $C=2,-1$ that are separated by a trivial phase with $C=0$ \cite{Li2016}. The transition between these phase occurs when the bulk-gap closes, which for fixed values of $JS$ and $\Delta$ yield the following critical chemical potentials \,\cite{Li2016}
\begin{align}
\label{eq:crit}
\mu_c &= \pm \sqrt{(JS)^2 - \Delta^2} \\ \nonumber
\mu_c &= \mp 4t \pm \sqrt{(JS)^2 - \Delta^2} \ .
\end{align}
It immediately follows from these criteria that spatial disorder in any of the parameters can locally tune the system between topological trivial and non-trivial phases. For the parameters used in Fig.~\ref{fig:Chern number}, one obtains $\mu_c/t=\pm 1.6, \pm 2.4$.

The phase diagrams Figs.~\ref{fig:Chern number} (a)-(d) reveal that the overall effect of all four types of disorder is similar in that the topological phases are suppressed with increasing disorder strength. However, the critical disorder strength at which the topological phase collapses depends on the chemical potential: the further the system is located from critical chemical potential, $\mu_c$, of the clean system, the larger is the critical disorder strength required to destroy the topological phase. The origin of this dependence lies in the fact that the gap protecting the topological phase increases with increasing distance from $\mu_c$, thus necessitating a larger disorder strength to close it and to drive the system trivial. There are, however, some noteworthy characteristics regarding the effects of the various types of disorder. In particular, we find that the topological phases are more robust against potential disorder than magnetic disorder in $J$ (the latter being described by $H_J^{(1)}$). Moreover, for the case when the spin orientation deviates from the $z$-axis, the topological phases are destroyed when the spin orientation is uniformly distributed over the upper hemisphere (i.e., for $\theta=\pi/2$). Finally, percolation possesses the strongest detrimental effect on the stability of the topological phases. For example, for $\mu=1.4t$, the topological phase is destroyed by magnetic disorder for $w_J = 2t$ (corresponding to half of the electronic bandwidth), while in the case of percolation, the topological phase is already destroyed for $p \approx 0.1$.

In Figs.~\ref{fig:Chern number} (e)-(h), we present line-cuts of the Chern number with increasing disorder strength for several values of $\mu$ [these lines correspond to vertical cuts in Figs.~\ref{fig:Chern number} (a)-(d)]. These line-cuts reveal that due to the finite size of the system, the disorder-induced transition between topological and non-topological phases is smooth and continuous, and thus represents a crossover (exhibiting a non-quantized Chern number), rather than a phase transition. However, a comparison of the Chern number line cuts for different system sizes [see dashed lines in Fig.~\ref{fig:Chern number}(e) which were computed for a $(50 \times 50)$ system] reveals that the transition becomes increasingly sharper and evolves toward a step-like function with increasing system size as expected for a phase transition. We therefore conclude that in the thermodynamic limit, a phase transition will occur at a critical value of the disorder strength separating a topological phase with a quantized Chern number, from a non-topological phase with $C=0$.

\begin{figure}[t]
\begin{center}
\includegraphics[width=8.5cm]{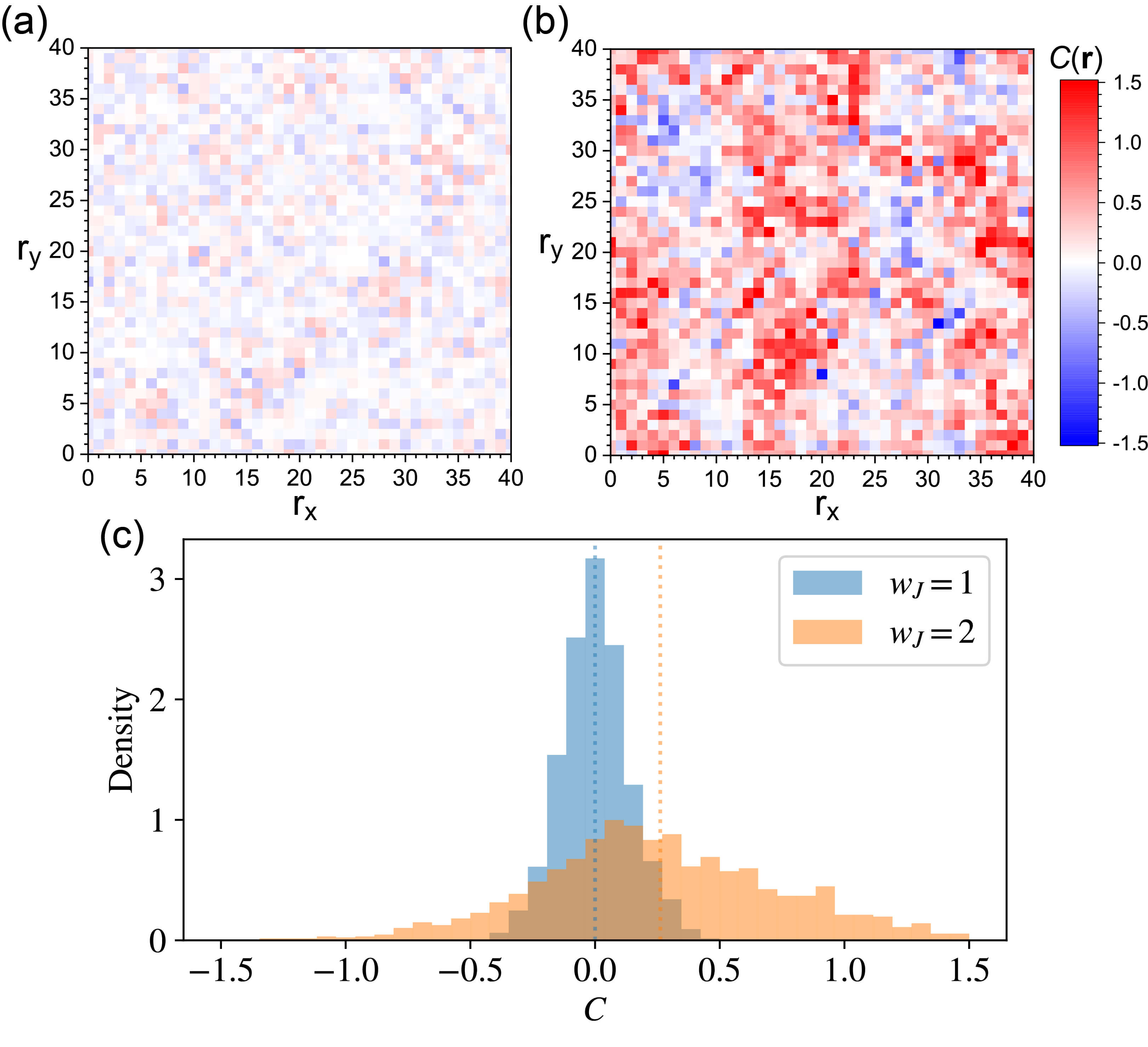}
\caption{Spatial plots of the Chern number density $C({\bf r})$ for the case of magnetic disorder with $\mu=-2t$ and (a) $w_J=t$, and (b) $w_J=2t$ for a $41 \times 41$ system. (c) Distributions of the Chern number density $C({\bf r})$ for the two case shown in panels (a),(b). The vertical dashed lines show the mean values of the distribution, corresponding to the macroscopic Chern numbers. The histogram is scaled such that the integral over the histogram is equal to unity. Parameters are $(JS,\alpha,\Delta)=(2.0,0.8,1.2)t$.}
\label{fig:mu2}
\end{center}
\end{figure}

It is interesting to note that for $\mu=-2t$ (which corresponds to the topological trivial phase in the clean limit) magnetic disorder locally induces domains of a topological phase, leading to a non-zero Chern number [see blue solid line in Fig.~\ref{fig:Chern number}(f)]. This can be understood as follows: according to Eq.(\ref{eq:crit}), local variations in $J$ will lead to local variations in $\mu_c$, which implies that even for $\mu=2t$, the system can be locally in a topological phase, if $J_{\bf r}S$ is sufficiently large. This is borne out by the plot of the Chern number density $C({\bf r})$, shown in Figs.~\ref{fig:mu2}(a),(b), for $w_J=t$ and $w_J=2t$, respectively. In particular, for $w_J=2t$, we have $C \approx 0.26$, and Fig.~\ref{fig:mu2}(a) shows the existence of larger domains of positive, non-zero $C({\bf r})$. In contrast, for $w_J=t$, with $C \approx 0.0$, Fig.~\ref{fig:mu2}(b) shows no discernible domains, but only a random distribution of small values of $C({\bf r})$. This difference is even more apparent when considering the distribution of $C({\bf r})$ for these two cases presented in Fig.~\ref{fig:mu2}(c). While the distribution of $C({\bf r})$ for $w_J=t$ is centered around zero, for $w_J=2t$ it is considerable broader and has shifted to positive values, resulting in a non-zero $C \approx 0.26$. Note that in the absence of any disorder and $C=0$, the Chern number density $C({\bf r})=0$ for all $\bf r$.

\section{Correlated Potential Disorder}

To understand how disorder leads to the collapse of topological phases, it is instructive to consider the spatial correlations between the local disorder, the particle number density, and the Chern number density. To this end, we consider a spatially correlated potential disorder that allows for the emergence of larger domains of nearly the same disorder potential, which facilitates the spatial comparison. As before we start by considering a random potential disorder, as described by Eq.(\ref{eq:pot}), but then replace $U_{\bf r}$ by the disorder ${\overline { U}}_{\bf r}$ which is generated from $ U_{\bf r}$ by using a low-pass filter via
\begin{align}
 {\overline { U}}_{\bf r} = \mathcal{F}_{\bf r}^{-1}[\mathcal{F_{\bf k}}[ U_{\bf r}] e^{-{\bf k}^2/{\bf K}_c^2}]
\label{eq:CorrDisorder}
\end{align}
where $\mathcal{F}$ is the Fourier transform over ${\bf r}$ and ${\bf K}_c$ is the cut-off wavevector. This low pass filter implies that the short-wavelength fluctuations in the disorder potential $ U_{\bf r}$ with wave-number $k>|{\bf K}_c|$ are eliminated, which smoothes the disorder potential and increases the disorder correlations as described by
\begin{align}
\rho_{\bf \delta} = \frac{\langle {\overline { U}_{\bf r}} {\overline { U}_{\bf r+\delta}} \rangle - \mu_U^2}{\sigma_U^2}
\end{align}
where the mean value $\mu_U = \langle {\overline { U}}_{\bf r} \rangle \equiv 0$, and the variance is $\sigma_U^2 = \langle \left( {\overline { U}}_{\bf r} \right)^2 \rangle $. For random disorder (corresponding to $|{\bf K}_c|=\infty$), we obtain for nearest neighbor correlations (i.e, ${\bm \delta}= {\hat x}, {\hat y}$)  $\rho_{\hat x}\leq 10^{-6}$ (which vanishes in the thermodynamic limit). In contrast, for the case $|{\bf K}_c|=\pi$, which we consider below as an example for correlated disorder, we have $\rho_{\hat x} \approx 0.04$. This implies that the disorder potential develops short range correlations with decreasing $|{\bf K}_c|$, as the systems begins to exhibit larger domains of the same potential.

\begin{figure}[t]
\begin{center}
\includegraphics[width=8.5cm]{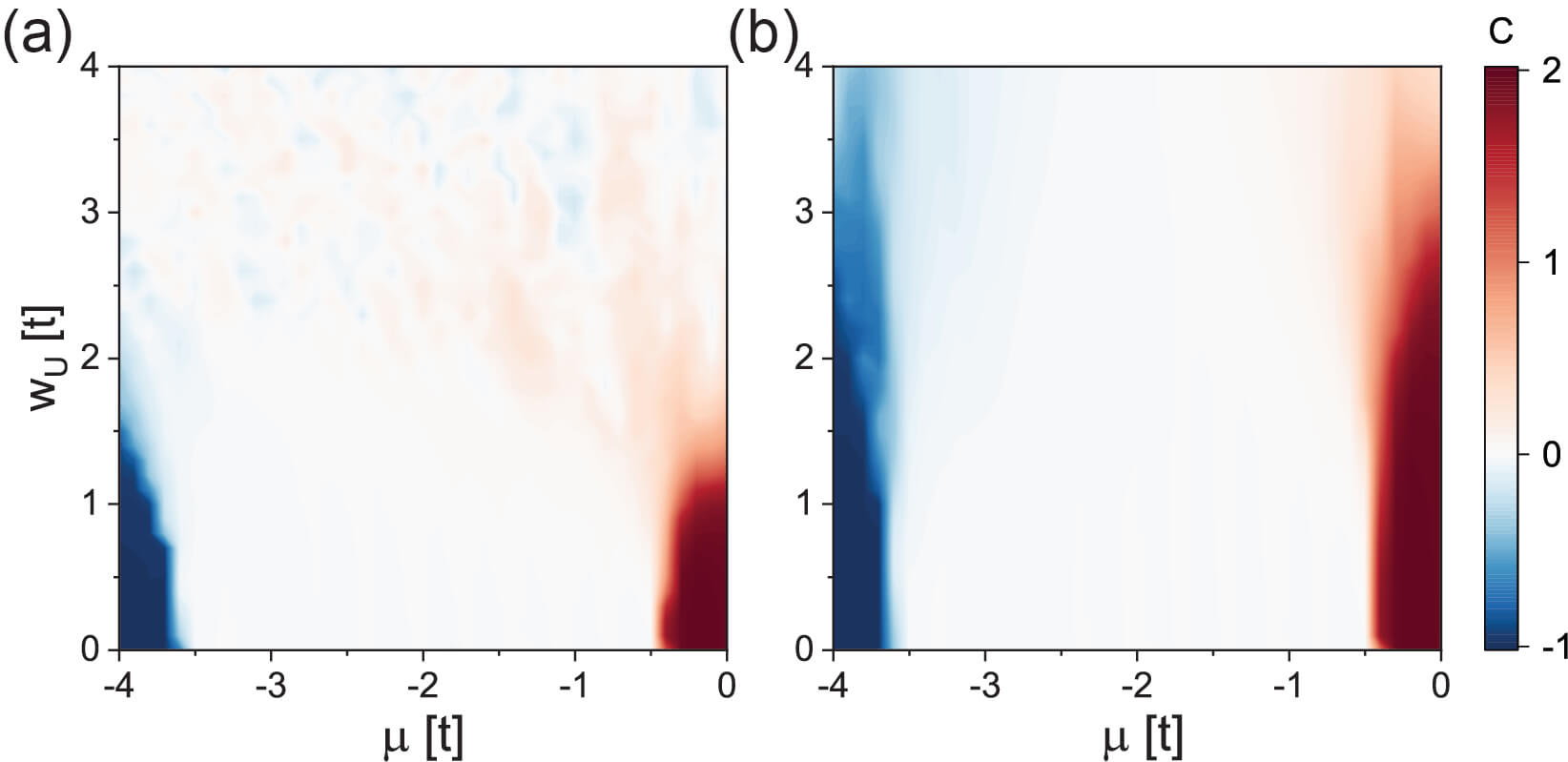}
\caption{Comparison of the topological phase diagram for (a) uncorrelated, and (b) correlated disorder [Eq.(\ref{eq:CorrDisorder})] with $|{\bf K}_c|=\pi/2$. The phase diagrams were obtained for a $(30 \times 30)$ system with parameters $(JS,\alpha,\Delta)=(0.5,0.2,0.3)t$, yielding $\mu_c/t=\pm 0.4, \pm 3.6$ in the clean case.}
\label{fig:CorrPhaseDiagram}
\end{center}
\end{figure}
In Figs.\ref{fig:CorrPhaseDiagram}(a),(b), we present the topological phase diagrams for random and correlated potential disorder, respectively, where the latter was obtained using Eq.(\ref{eq:CorrDisorder}) with $|{\bf K}_c|=\pi$.  To demonstrate the generality of our results, we consider a set of parameters, $(JS,\alpha,\Delta)=(0.5,0.2,0.3)t$, yielding $\mu_c/t = \pm 0.4,  \pm 3.6$ in the clean case, that is different from that employed in Fig.~\ref{fig:Chern number}.  A comparison of these two phase diagrams reveals that correlated disorder possesses a weaker effect on the stability of the topological phases than random potential disorder, requiring thus a larger critical disorder strength to destroy the topological phases. We note that the effect of correlated potential disorder weakens with decreasing $|{\bf K}_c|$.

\begin{figure}[t]
\begin{center}
\includegraphics[width=8.5cm]{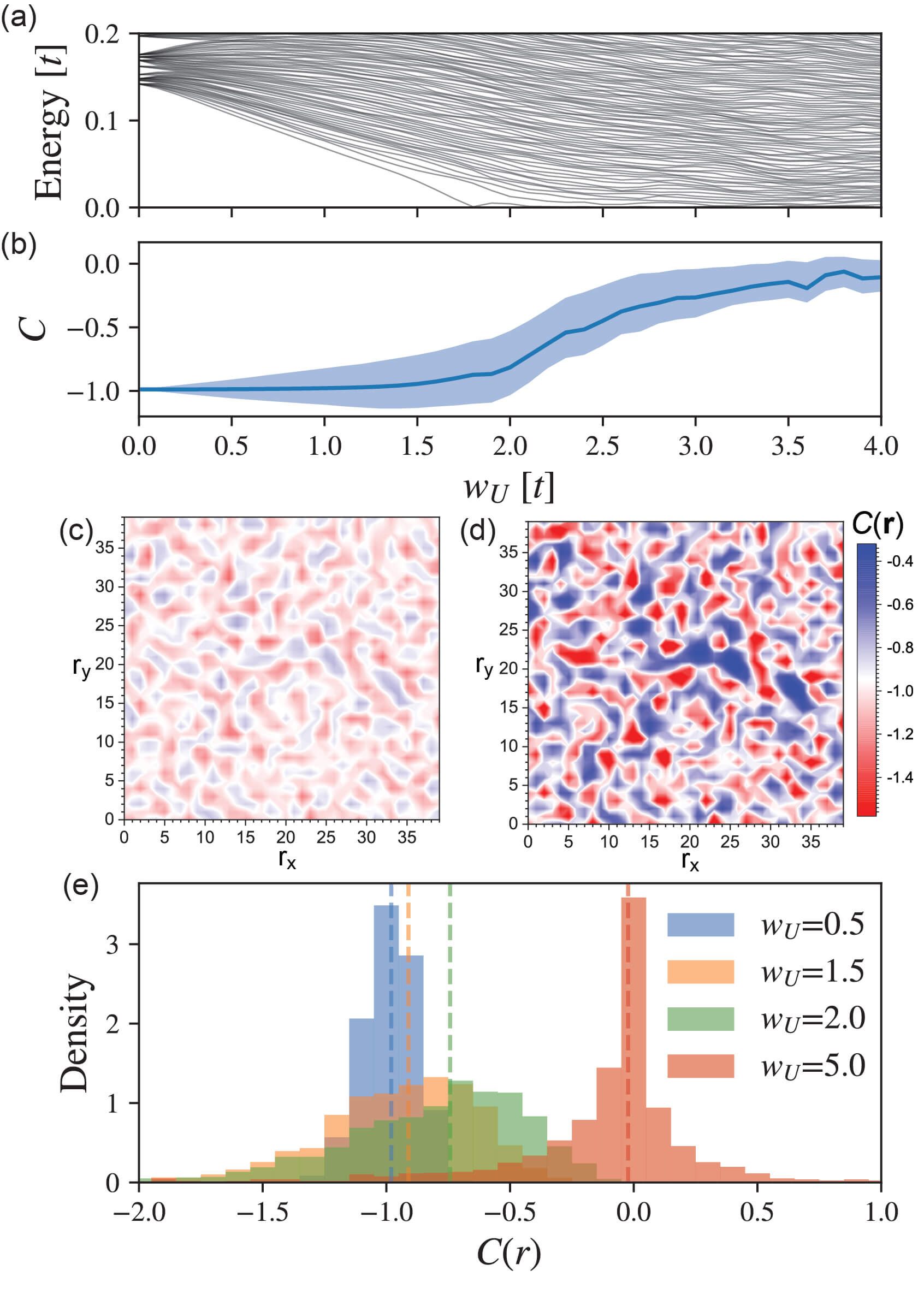}
\caption{Evolution of (a) the lowest energy eigenstates, and (b) the Chern number and the upper and lower quartiles (light blue area) of the Chern number density with increasing disorder strength $w_U$ for $|{\bf K}_c|=\pi$ and $\mu=-4t$, corresponding to the $C=-1$ phase in the clean limit. The critical disorder value is given by $w^c_U \approx 1.8t$ for this particular disorder realization.  Spatial plot of the Chern number density $C(r)$ for (c) $w_U=0.5t$, and (d) $w_U=1.5t $. (e) Distributions of the Chern number density $C({\bf r})$ for different values of disorder strength $w_U$. The vertical dashed lines show the mean values of the distribution, corresponding to the macroscopic Chern numbers. These results were obtained for a $(40 \times 40)$ system with parameters $(JS,\alpha,\Delta)=(0.5,0.2,0.3)t$.}
\label{fig:Q1}
\end{center}
\end{figure}

In Figs.~\ref{fig:Q1}(a),(b) we present the evolution of the lowest energy eigenstates and the Chern number with increasing $w_U$ for correlated potential disorder with $|{\bf K}_c|=\pi$.
The critical disorder strength where the first eigenstate reaches zero energy is given by $w^c_U \approx 1.8t$. $w^c_U$ varies between different disorder realizations, and possesses a disorder-averaged value (using $n=50$ disorder realizations) of $\langle w^c_U \rangle \approx 2.0$. While the Chern number is not quantized any longer for any finite disorder strength, it is substantially reduced from its value $C=-1$ in the clean system only for $w_U>w^c_U$. To understand how this departure from $C=-1$ occurs, we present in Figs.~\ref{fig:Q1}(c) and (d) a spatial plot of $C({\bf r})$ for $w_U=0.5t$ and $w_U=1.5t$, respectively; for both values $w_U<w^c_U$.  As expected, we find that disorder results in an inhomogeneous spatial form of $C({\bf r})$ and that with increasing disorder strength, the spatial variations in $C({\bf r})$ increase as well. This can be nicely visualized by plotting a histogram of $C({\bf r})$ [see Fig.~\ref{fig:Q1}(e)] which shows that increasing the disorder strength leads to a broadening of the $C({\bf r})$ distribution. However, only for $w_U>w^c_U$ does the entire distribution shift to lower values [see $w_U=2.0t$ in Fig.~\ref{fig:Q1}(e)], resulting in a decrease of the Chern number. For $w_U \gg w^c_U$ [see $w_U=5.0t$ in Fig.~\ref{fig:Q1}(e)], the distribution becomes centered around zero, leading to a vanishing Chern number.

\begin{figure}[t]
\begin{center}
\includegraphics[width=8.5cm]{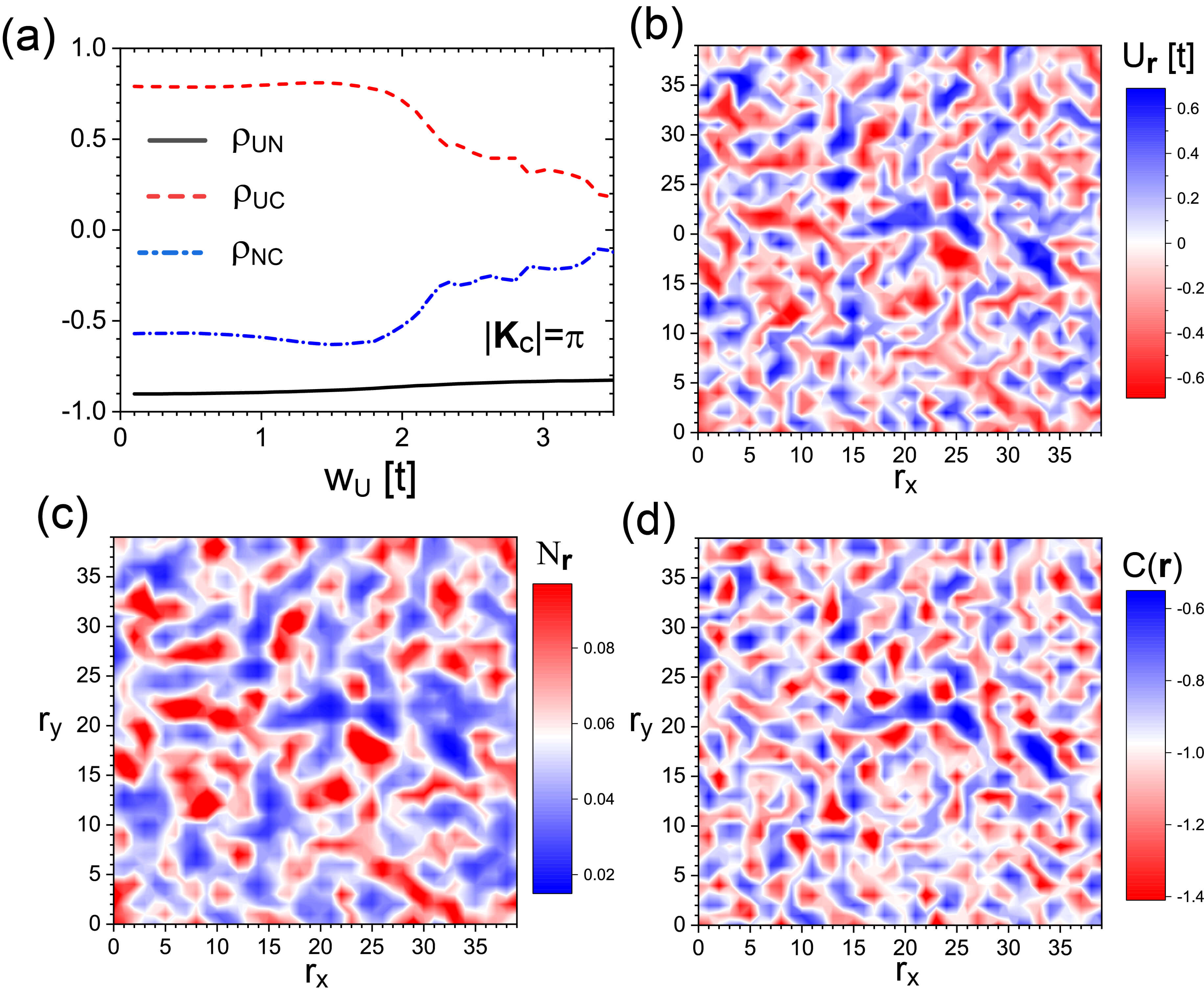}
\caption{(a) Pearson correlation $\rho_{X,Y}$ for the disorder potential, ${\overline {U}}_{\bf r}$ (denoted by $U$), the particle density $N_{\bf r}$ (denoted by $N$), and the Chern number density $C({\bf r})$ (denoted by C). Spatial plot of (b) ${\overline { U}}_{\bf r}$, (c) $N_{\bf r}$, and (d) $C({\bf r})$ for $w_U = 1.5t$}
\label{fig:Q1_corr}
\end{center}
\end{figure}
To investigate the correlations of the spatial structure of $C({\bf r})$ with other physical observables in the system, we consider the Pearson correlation function between two physical observables $X({\bf r})$ and $Y({\bf r})$ defined via
\begin{align}
\rho_{X,Y}= \left\langle \frac{X({\bf r})-\mu_X}{\sigma_X} \frac{Y({\bf r})-\mu_Y}{\sigma_Y}  \right\rangle
\end{align}
where $\mu_i,\sigma_i$ $(i=X,Y)$ are the expectation value and standard deviation of the observable $i$. In Fig.~\ref{fig:Q1_corr}(a), we present the correlation functions for the disorder potential, ${\overline { U}}_{\bf r}$, the particle density $N_{\bf r}$, and the Chern number density $C({\bf r})$. As expected, we find that $ {\overline {U}}_{\bf r}$ and $N_{\bf r}$ are nearly completely anti-correlated, with a local increase in $ {\overline { U}}_{\bf r}$ (i.e., creating a repulsive potential) leading to a decrease in $N_{\bf r}$. Interestingly enough, we find that there also exists a substantial correlation between $C({\bf r})$ and $ {\overline {U}}_{\bf r}$, and thus also between $C({\bf r})$ and $N_{\bf r}$. This is also evident from a comparison of the spatial form of $ {\overline { U}}_{\bf r}$, $N_{\bf r}$ and $C({\bf r})$,  shown in Figs.~\ref{fig:Q1_corr}(b)-(d) for the case of $w_U=1.5t$. All three properties possess to a large extent the same spatial structure, in agreement with the substantial correlation revealed by $\rho_{X,Y}$ shown in Fig.~\ref{fig:Q1_corr}(a). This result might open a new approach to investigating the form of the Chern number density, and hence the Chern number, in real space, through measurements, for example, of the particle number density.

\section{Conclusions}

We have investigated the effects of various types of potential and magnetic disorder on the stability of topological superconductivity in two-dimensional magnet superconductor hybrid systems. These hybrid structures are of great current interest as they represent a promising platform for engineering Majorana fermions. We showed that random potential disorder leads to the weakest, while percolation disorder leads to the strongest suppression of topological superconducting phases. Moreover, random magnetic disorder can lead to the formation of local topological domains, even if the bulk system is in a topologically trivial phase. We also demonstrated that spatially correlated potential disorder exerts a weaker effect on the topological phase diagram than random disorder. Moreover, we showed that disorder leads to a spatially inhomogeneous form of the Chern number density, the width of whose distribution increases with increasing disorder. We also demonstrated that the disorder induced phase transition from topological to trivial phases is accompanied by a downward shift of the distribution of the Chern number density, becoming centered around zero, and leading to a vanishing mean, i.e., macroscopic Chern number. However, even in the topological trivial phase, spatial domains of non-zero Chern number density remain. Finally, we showed that there exist considerable spatial correlations between the spatial structure of the Chern number density, the potential disorder, and the particle number density. This result might open a new approach to detecting the Chern number density, and hence the Chern number, in real space through measurements of the particle density. \\

\acknowledgements

This work was supported by the U. S. Department of Energy, Office of Science, Basic Energy Sciences, under Award No. DE-FG02-05ER46225 (EM,CA, and DKM). SR acknowledges support from an Australian Research Council Future Fellowship (FT180100211).


%

\end{document}